# Branch-counting in the Everett interpretation of quantum mechanics[†]

Simon Saunders

Abstract: A defence is offered of a version of the branch-counting rule for probability in the Everett interpretation (otherwise known as many-worlds interpretation) of quantum mechanics that both depends on the state and is continuous in the norm topology on Hilbert space. The well-known branch-counting rule, for realistic models of measurements, in which branches are defined by decoherence theory, fails this test. The new rule hinges on the use of decoherence theory in defining branching structure, and specifically decoherent histories theory. On this basis ratios of branch numbers are defined, free of any convention. They agree with the Born rule, and deliver a notion of objective probability similar to naïve frequentism, save that the frequencies of outcomes are not confined to a single world at different times, but spread over worlds at a single time. Nor is it ad hoc: it is recognizably akin to the combinatorial approach to thermodynamic probability, as introduced by Boltzmann in 1879. It is identical to the procedure followed by Planck, Bose, Einstein and Dirac in defining the equilibrium distribution of the Bose-Einstein gas. It also connects in a simple way with the decision-theory approach to quantum probability.

1.Introduction

The unique appeal of Everett's approach to the foundational questions of quantum mechanics is that whilst it is demonstrably a form of realism, on taking the quantum state to represent something existent, it leaves quantum mechanics unchanged -- save in this one respect: the Schrödinger equation is taken to apply without restriction, and in particular to apply to macroscopic bodies. Everett showed that on this basis, when a sequence of measurements is performed, the result is a superposition of states, what Everett called *branches*, each of which describes some definite statistics of outcomes. In this way he was able to show how the deterministic Schrödinger equation could be reconciled with indeterminism and statistical data. The implication, however, if the quantum state is or represents something existent, is fantastical: all the outcomes in a quantum measurement actually exist, including the experimenters witnessing them. This is many worlds: according to the Everett interpretation of quantum mechanics, there is a vast multiplicity of worlds, constantly branching in time, containing people just like you and me.

Extraordinary claims require extraordinary arguments; the bar, rightfully, should be high. But Everett was less than clear in two respects. First, the quantum state can equally be written as a superposition of any set of basis states; what reason is there to single out his 'branch states' as distinguished? This is 'the preferred basis problem'. Second, Everett assumed that a probability measure over branches is a function of branch amplitude[1] and phase, and from this derived the Born rule (the standard probability rule for quantum mechanics). But there is a natural alternative probability measure suggested exactly by his picture of branching, that is not of this form and that does *not* in general agree with the Born rule: the 'branch-counting rule'. Let the world split into two, each with a different amplitude: in what sense can one branch be more probable than the other? If Everett is to be believed, both come into existence with certainty, regardless of their amplitudes. After many splittings of this kind, divide the number of branches with one outcome, by the number of branches with another; that should give their relative probability. This is the branch-counting rule, in general in contradiction with the Born rule.

---

[1] By the amplitude of a state I mean its Hilbert-space norm, a non-negative real number. In what follows, states written in Dirac notation $|\varphi\rangle \in \mathcal{H}$ are not assumed to be normalized to unity (in fact only ratios in norms ever matter).

Everett did not reply to either criticism, having left the field even before the publication, in *Reviews of Modern Physics* in 1957, of his doctoral thesis, all of eight pages in length; he never wrote on quantum mechanics again. Bryce De Witt, Everett's main champion in the early years, was likewise silent on these matters.[2] But others since have replied to both, and on the point of the preferred basis problem, in constructive detail. However, when it comes to the branch-counting rule, it has only been *discredited* -- on the grounds that in realistic models of measurements, no meaningful notion of branch number is to be found.[3] That raises a worry all on its own: how can there be large numbers of branches (or 'worlds'), according to unitary quantum mechanics, if the notion of branch number makes no sense?

There is unfinished business with the branch-counting rule in the Everett interpretation. This is our main topic, but it is connected to the preferred basis problem, and cannot really be considered independent of it. By wide consensus, the latter was substantially resolved by appeal to decoherence theory, as developed in the '80s and '90s. But agreement, where broad, is often shallow, and the details of why, exactly, decoherence theory is important to the Everett interpretation are still disputed. Those details matter to the branch-counting rule, and we shall have to revisit them.

The argument that follows is that whilst the original rule is indeed inapplicable, when branches are defined in decoherence theory, there is another branch-counting rule which does still make sense. The new rule, call it the 'equi-amplitude rule', is well-defined and, for ratios in branch numbers, free of convention. Unlike the old method of counting, those ratios agree with the Born rule. Nor is it ad hoc: it is modelled on Boltzmann's method of counting microstates, used to great effect in the early days of statistical mechanics. The analogous procedure, using the quantum histories formalism, gives the new rule, with states of quasiclassical histories (in the sense of Gell-Mann and Hartle (1989)) as branches.[4]

We begin in §2 with Everett's concept of branching, and in the next section, with its relation to decoherence theory. The implications for the old branch-counting rule are considered in §4. That identifies a condition of adequacy for any branch-counting rule, that it should depend continuously on the state. Boltzmann's state-counting rule is introduced in §5, and the basics of decoherent histories theory, needed to apply it to branching, in §6. There we conclude that histories of a quasi-classical domain, in Gell-Mann and Hartle's sense, are branches, in Everett's sense. The new branch-counting rule, modelled on Boltzmann's rule, follows immediately. It is considered n §7 with respect to its formal properties, and in §8, in connection with wider philosophical questions, including its relation to the decision-theory approach to quantum probability (Deutsch 1999, Wallace 2012). In the concluding section 9 we return to the question of whether the new rule really is the same as Boltzmann's method -- in particular as it was applied to black-body radiation in the early days of quantum mechanics. As we shall see, so far from being ad hoc, Planck, Bose, Einstein and Dirac all used the new branch-counting rule, but in the special case of a completely degenerate equilibrium state.

---

[2] The nearest he came to comment on the basis problem was in a footnote (DeWitt 1971 p.210) 'only those decompositions are meaningful which reflect the behavior of a concrete dynamical system'; true enough, but rather brief.

[3] Saunders (2005 p.25), Wallace (2012 p.99-102). De Witt's student Neil Graham (who was the first to state the branch-counting rule) suggested an alternative, in agreement with the Born rule (Graham 1973), but it was clearly ad hoc.

[4] It can also be applied to pilot-wave theory, with equi-amplitude coarse-grainings of Bohmian trajectories replacing equi-amplitude quasiclassical histories, with the difference that these are numbers of possible histories, rather than existing histories. Here I consider only the Everett interpretation.

## 2. Everett's model of measurement

Consider a measurement of the z-component of spin, as in the Stern-Gerlach experiment. Suppose that the measurement is repeatable, so that the same spin-system can be measured again, yielding the same outcome. The unitary evolution should then satisfy the protocols:

$$|\varphi_\uparrow\rangle|\Phi_0\rangle \to |\varphi_\uparrow\rangle|\Phi_\uparrow\rangle \qquad (2.1)$$
$$|\varphi_\downarrow\rangle|\Phi_0\rangle \to |\varphi_\downarrow\rangle|\Phi_\downarrow\rangle$$

where $|\Phi_0\rangle, \Phi_\uparrow\rangle, \Phi_\downarrow\rangle$ are states of the macroscopic apparatus (reading 'ready', 'spin-up', 'spin-down', respectively). Following Everett, we also include a memory registrar, with states $|\Psi_s\rangle$, where s is a 'memory sequence' – a sequence of arrows ↑ and ↓, the results of previous spin-measurements. Allowing for a 'reset' operation the unitary evolution of the display and memory should satisfy:

$$|\Phi_\uparrow\rangle|\Psi_s\rangle \to |\Phi_\uparrow\rangle|\Psi_{s\uparrow}\rangle \to |\Phi_0\rangle|\Psi_{s\uparrow}\rangle$$
$$|\Phi_\downarrow\rangle|\Psi_s\rangle \to |\Phi_\downarrow\rangle|\Psi_{s\downarrow}\rangle \to |\Phi_0\rangle|\Psi_{s\downarrow}\rangle.$$

These, like the previous protocols, are fully deterministic, so can be implemented by a Schrödinger equation without any probability interpretation. The complete cycles for these two cases are:

$$|\varphi_\uparrow\rangle|\Phi_0\rangle|\Psi_s\rangle \to |\varphi_\uparrow\rangle|\Phi_\uparrow\rangle|\Psi_s\rangle \to |\varphi_\uparrow\rangle|\Phi_\uparrow\rangle|\Psi_{s\uparrow}\rangle \to |\varphi_\uparrow\rangle|\Phi_0\rangle|\Psi_{s\uparrow}\rangle \qquad (2.2)$$
$$|\varphi_\downarrow\rangle|\Phi_0\rangle|\Psi_s\rangle \to |\varphi_\downarrow\rangle|\Phi_\downarrow\rangle|\Psi_s\rangle \to |\varphi_\downarrow\rangle|\Phi_\downarrow\rangle|\Psi_{s\downarrow}\rangle \to |\varphi_\downarrow\rangle|\Phi_0\rangle|\Psi_{s\downarrow}\rangle.$$

It then follows from the orthogonality conditions and from the linearity of the Schrödinger equation that if the measured system is initially in a superposition of spin eigenstates

$$|\varphi\rangle = a|\varphi_\uparrow\rangle + b|\varphi_\downarrow\rangle \qquad (2.3)$$

where $a$ and $b$ are non-zero complex numbers, the evolution is

$$|\varphi\rangle|\Phi_0\rangle|\Psi_s\rangle$$
$$\to a|\varphi_\uparrow\rangle|\Phi_\uparrow\rangle|\Psi_s\rangle + b|\varphi_\downarrow\rangle|\Phi_\downarrow\rangle|\Psi_s\rangle$$
$$\to a|\varphi_\uparrow\rangle|\Phi_\uparrow\rangle|\Psi_{s\uparrow}\rangle + b|\varphi_\downarrow\rangle|\Phi_\downarrow\rangle|\Psi_{s\downarrow}\rangle$$
$$\to a|\varphi_\uparrow\rangle|\Phi_0\rangle|\Psi_{s\uparrow}\rangle + b|\varphi_\downarrow\rangle|\Phi_0\rangle|\Psi_{s\downarrow}\rangle.$$

The conventional reading is that in this final state no definite measurement is recorded in memory, because it *has* no definite state (it is in an entangled state); the reading Everett gave is that the evolution is a superposition of two *sequences*, in each of which the apparatus, or rather *an* apparatus (note the language difficulty), goes through two distinct cycles, each giving a different outcome, each with a different memory.

To see the interpretation in action again, let a second measurement be made on the same spin-system as before (recall we assumed the experiment was repeatable). The unitary evolution is now (with subscripts to indicate the first and second measurements):

$$(a|\varphi_\uparrow\rangle + b|\varphi_\downarrow\rangle)|\Phi_0\rangle|\Psi_s\rangle$$
$$\xrightarrow{1} a|\varphi_\uparrow\rangle|\Phi_\uparrow\rangle|\Psi_s\rangle \;\; + b|\varphi_\downarrow\rangle|\Phi_\downarrow\rangle|\Psi_s\rangle$$
$$\xrightarrow{1} a|\varphi_\uparrow\rangle|\Phi_\uparrow\rangle|\Psi_{s\uparrow}\rangle \;\; + b|\varphi_\downarrow\rangle|\Phi_\downarrow\rangle|\Psi_{s\downarrow}\rangle$$
$$\xrightarrow{1} a|\varphi_\uparrow\rangle|\Phi_0\rangle|\Psi_{s\uparrow}\rangle \;\; + b|\varphi_\downarrow\rangle|\Phi_0\rangle|\Psi_{s\downarrow}\rangle$$
$$\xrightarrow{2} a|\varphi_\uparrow\rangle|\Phi_\uparrow\rangle|\Psi_{s\uparrow}\rangle \;\; + b|\varphi_\downarrow\rangle|\Phi_\downarrow\rangle|\Psi_{s\downarrow}\rangle \tag{2.4}$$
$$\xrightarrow{2} a|\varphi_\uparrow\rangle|\Phi_\uparrow\rangle|\Psi_{s\uparrow\uparrow}\rangle \;\; + b|\varphi_\downarrow\rangle|\Phi_\downarrow\rangle|\Psi_{s\downarrow\downarrow}\rangle$$
$$\xrightarrow{2} a|\varphi_\uparrow\rangle|\Phi_0\rangle|\Psi_{s\uparrow\uparrow}\rangle \;\; + b|\varphi_\downarrow\rangle|\Phi_0\rangle|\Psi_{s\downarrow\downarrow}\rangle.$$

Applying the same 'vertical' reading of the state, we see there are again two sequences of states, neither satisfying the Schrödinger equation on its own, but doing so when superposed. Each sequence describes a definite outcome on the first measurement, and on the second measurement, each describes the *same* outcome as resulting – in the one case $|\Psi_{s\uparrow\uparrow}\rangle$, in the other $|\Psi_{s\downarrow\downarrow}\rangle$. In this way Everett was able to recover an 'effective' projection postulate, along branches, following from the Schrödinger equation.

The other major part of the measurement postulates, the Born rule, was recovered in terms of a measure over branches: Everett showed that the Born rule is the only additive measure over orthogonal branch states that is a function of amplitude and phase. He also proved a quantum version of the law of large numbers: with high probability (squared amplitude), the recorded statistics in each branch are in approximate agreement with the Born rule. In the limiting case of infinitely many trials, only branches exactly agreeing with the Born rule remain.

In this, what Everett called the 'abstract' theory of measurement, the basis used in decomposing the total state – into states describing the apparatus and memory as having definite readings and records -- was simply postulated. The problem of how to improve on this is the preferred basis problem – why privilege those states? My reason for revisiting Everett's treatment is to highlight that even after the basis states are (somehow) defined, there still remains what might be called the preferred *branch* problem – the problem of how, given a basis, states are collected into sequences, into branches. Sums of states, as in (2.4), can be written in any order: what dictates the correct 'vertical' reading?

Everett made no comment to this matter; what he did was use the measurement protocols, equations (2.2), and in particular the mechanism of memory: the last-most memory state determines all the others leading up to it. But those protocols, like the basis states, were just *stipulated*. To many this reasoning seemed circular.[5] Yet it was a virtuous circle: the preferred basis, and the preferred branches, come together; the unitarily evolving state has a branching structure, insofar as with respect to some basis, a vertical reading can be found, in accordance with definite rules, as follows from the Schrödinger equation. Whether such a reading can be given is highly non-trivial as soon as we go beyond the abstractions of measurement protocols, even when the initial state and dynamics involves a physical artifact (the apparatus) of our own making. In seeking a comparable interpretation of the state for natural processes, absent human intervention, the impression that it can be conjured up at will vanishes altogether.

---

[5] This was Bell's main criticism of the Everett interpretation (Bell 1987 p.98, p.192). (Another criticism was that branching involves an arrow of time, absent in the unitary formalism; a point I come back to in §6.)

This is the core of the Everett interpretation. Note again how it contradicts the standard interpretation of a state like (2.4), when it arises at the microscopic level (quantum mechanics is not standardly applied to macroscopic bodies): the standard reading is that in such a case, in an entangled state, no definite state can be attributed to the subsystems involved. To counter this, Everett introduced his notion of 'relative state', applicable at the micro- and macro-level, whatever the state. That, in retrospect, was a tactical mistake, distracting from the interpretative method that he actually followed: a vertical reading of the state in accordance with rules. This is hardly ever possible for entangled microscopic systems. Moreover, a little reflection shows that these 'vertical' readings are entirely uncontroversial in other, classical, wave theories. Thus, to use an example given in Wallace (2013 p.464), we say that there are *two* beams of light, in the electromagnetic field, pointing in two different directions, in a superposition; we do not say there is a single beam of light of indeterminate direction, in a superposition of two states of the electromagnetic field, in one state pointing one way, in the other state pointing the other; we do not say there is no beam of light at all, because a beam of light is something with a definite direction, and the beam of light in this case does not have a definite direction. The 'beam states' and rules (rectilinear propagation) come together, following from the initial state and Maxwell equations, possibly involving the very same degrees of freedom (the same frequencies) of the electromagnetic field.

Evidently examples like this can easily be multiplied. There are two voices, or two melodies, in a superposition, two circular ripples in water, superposed; and so on. Again (returning to the 'language difficulty'), it is not the thing (beam, voice, melody, ripple) that is in a superposition, but degrees of freedom that are in superpositions. Speaking the same way, it is not that the memory registrar is in a superposition, it is that the registrar degrees of freedom are in a superposition.

Returning to the preferred basis problem, the symmetries of Hilbert space dictate that any basis can be used, for any Hamiltonian, any degrees of freedom, in any initial state, for representing a solution to the Schrödinger equation; why represent it in one basis rather than another? Or is the state expressed in every basis a distinct multiplicity, all them existences? But whether or not there is a basis satisfying definite rules *does* depend on the basis, Hamiltonian, degrees of freedom, and initial state. There is no temptation, in interpreting the evolving state of the electromagnetic field in terms of two beams of light, to reify the countless other ways of representing the state of the field at each instant in time. Neglect of the preferred branch problem made the preferred basis problem seem much harder than it was, and the role of decoherence theory less transparent.[6]

---

[6] Indeed, some advocates of the Everett interpretation see little or no need for decoherence theory (Deutsch (1997, 2016) Vaidman (1998, 2019, 2021).

## 3. Decoherence theory as the general theory of branching

The upshot of this stage-setting is that Everett's branches are defined by a basis satisfying definite rules, as follow from the Schrödinger equation and initial state. The preferred basis problem (which basis?) and preferred branch problem (which rules?) are always solved together, and depend on the Hamiltonian and initial state. In Everett (1957), the basis and rules were abstractly defined, but in a longer draft (eventually published in De Witt and Graham (1973)) he sketched out a concrete model: localised states of centre of mass degrees of freedom, for large masses, with well-defined velocities (coherent states); if the gradient of the potential is approximately constant over the widths of those states, their motions will satisfy classical equations. The fact that over sufficiently long times, superpositions of such states would eventually be produced, is not a difficulty, for that is just more branching (Everett 1973 p.89-90).

Seen in this light, the obvious question is whether there may be *other* examples of rules and states coming together, as driven by unitary evolutions. Early work by Dieter Zeh and Wojciech Zurek in decoherence theory (to use the term introduced by Murray Gell-Mann in 1989) can be read in just this way. Subsequent landmark papers (Caldeira and Leggett 1983, Joos and Zeh 1984, Gallis and Fleming 1990, Tegmark 1993) barely mentioned the concept of branching, but showed that in scattering environments (gasses and radiation baths), coherent states for much smaller masses, going down to large molecules, obey quasiclassical equations ('quasiclassical' as going beyond classical Hamiltonian mechanics – as including friction, for example). But then they are branches in our sense.[7] It follows that a delocalised state of masses of the order of large molecules (and anything greater), in a scattering environment involving much lighter degrees of freedom, branches into localised states, each evolving quasiclassically, unable to interfere with each other. The suppression of interference is moreover extremely rapid; in gases at ordinary temperatures, it is much faster than thermal relaxation times.[8]

These examples are important because thermal environments are ubiquitous; for a tiny speck of dust, and anything larger, even the cosmic microwave background will do. They are called examples of 'quantum Brownian motion', because the term is linear in the velocity. There are many other examples of other kinds (see, for example, Stamp (2006)), but they fit the same mould: the Everett interpretation does not just concern the macroscopic, it concerns the emergence of dynamics at near-microscopic scales as well. And in quantum Brownian motion, molecular degrees of freedom do not only couple to the environment, they couple to each other -- whether by collisions, near-neighbour interactions, or exchanges of phonons. Superpositions of localised states will be produced all the time, for example, of reflected and transmitted states, produced in collisions, which since differing in velocities, rapidly decohere, hence produce branching; no special Hamiltonian (as in chaos) is needed. The implication is that for macroscopic numbers of degrees of freedom at ordinary temperatures, branching in Everett's sense is enormous, rapid, and ubiquitous.

---

[7] If more argument is needed, I refer to Kiefer (2002) for a survey of decoherence theory (in the sense of quantum Brownian motion) from the point of the decoherent histories formalism. He concludes with the observation (2002 p.257) that the latter language is much weaker, and can always be used. It can be used to express Everett's concept of branch as well, as I argue in §6. It follows that quantum Brownian motion can be interpreted in terms of branching.

[8] The coherence length $l$ is the maximum distance at which interference between localised states is possible. In scattering environments, it falls off as $1/\sqrt{\Lambda t}$, where $\Lambda$ is the localisation rate, until it reaches the de Broglie wavelength $h/\sqrt{mkT}$, at which it stabilises. This sets the widths of the preferred basis states. The localisation rate for a heavy molecule in air at ordinary temperature is of the order $\Lambda \sim 10^{34} m^{-2} s^{-1}$; for a particle of dust in the cosmic microwave background, $\Lambda \sim 10^{10} m^{-2} s^{-1}$ (see Joos 2002, p.83, p.63).

## 4. Branch counting and continuity

It thus follows from decoherence theory that huge numbers of branches are produced in a single run of any realistic experiment. For *each* outcome there are large numbers of branches, differing in ways irrelevant to the outcome. Why then count the relative numbers of *outcomes* – the *kinds* of branches? If existence is what matters, and all the branches exist, then it is the relative numbers of branches of each kind that matter, not the numbers of kinds.

With this correction, consider again the measurement of the z-component of spin in a Stern-Gerlach apparatus. We need another modification: there is no reason to suppose that exactly the same superposition is produced on each run of the apparatus. In replacing equation (2.1) by a more realistic model of measurement, we should consider only a single trial, rather than a protocol that applies to a large number of trials. If the initial state of the apparatus on that one run is $|\Phi_0\rangle$ (where we no longer bother to distinguish display and memory degrees of freedom), and if the initial state of spin is $|\varphi_\uparrow\rangle$, let there be $n_\uparrow$ orthogonal states $|\Phi_\uparrow^k\rangle$, $k = 1, \ldots, n_\uparrow$ produced, entering into the superposition, each describing an apparatus recording spin-up. Similarly, if the initial spin state is $|\varphi_\downarrow\rangle$, for that one run, let there be $n_\downarrow$ orthogonal states $|\Phi_\downarrow^k\rangle$, $k = 1, \ldots, n_\downarrow$ produced in a superposition, each describing an apparatus recording spin-down. Then for that one run we have in place of Eq(2.1):

$$|\varphi_\uparrow\rangle|\Phi_0\rangle \to |\varphi_\uparrow\rangle|\Phi_\uparrow^1\rangle + |\varphi_\uparrow\rangle|\Phi_\uparrow^2\rangle + \cdots + |\varphi_\uparrow\rangle|\Phi_\uparrow^{n_\uparrow}\rangle \qquad (4.1)$$

$$|\varphi_\downarrow\rangle|\Phi_0\rangle \to |\varphi_\downarrow\rangle|\Phi_\downarrow^1\rangle + |\varphi_\downarrow\rangle|\Phi_\downarrow^2\rangle + \cdots + |\varphi_\downarrow\rangle|\Phi_\downarrow^{n_\downarrow}\rangle.$$

But then the numbers would be the same were the initial state a superposition of spin-up and spin-down states, with any non-zero amplitudes $|a|$, $|b|$.[9] That is, the branch-counting rule would yield the same result (the same ratio $n_\downarrow/n_\uparrow$) whatever the initial state, on that single trial, so long as $a$ and $b$ are non-zero.

This independence of the branch-counting rule from the values of $a$ and $b$ (so long as non-zero) was already obvious in the abstract model, equation (2.4); new is that in practise all the action concerns details on how the apparatus responds to an initial state $|\varphi_\uparrow\rangle$, and an initial state $\varphi_\downarrow\rangle$, and is likely to vary from one run to the next. But further, if we are realistic about state *preparation* as well, we should recognise that in practise *the initial state of the spin system is never an exact eigenstate of the z-component of spin*. The 'z-component of spin' – any component of spin -- only has meaning insofar as a state-preparation device has a specified orientation relative to the measurement device; and relative orientation, even in quantum mechanics, varies continuously. (In the Bloch sphere of spin-states, in the norm topology, every vector has measure zero.) So since in reality, the state of the spin system is never an exact eigenstate of the z-component of spin, the ratio of branch numbers never depends on the initial state of the spin-system at all.[10] The rule *seemed* to interpret the state resulting from repeated measurements in terms of probability, but it did so only in a singular limit that is never actually attained. It is an artifact of an idealisation.

---

[9] This is only true for exact orthogonality of the decoherence basis, a condition that we shall later relax (this does not rescue the branch-counting rule now under discussion, however).

It will be helpful, for future reference, to formalise this difficulty. It is that the branch-counting rule is a discontinuous function of the state. Thus consider a sequence of initial spin states $|\varphi_k\rangle \in \mathcal{H}, , k = 1, 2, \ldots$ of the form

$$|\varphi_k\rangle = \cos\theta_k \, |\varphi_\uparrow\rangle + i \sin\theta_k \, |\varphi_\downarrow\rangle; \quad \theta_k = \frac{\pi}{2k+1}.$$

The $|\varphi_k\rangle$'s converge smoothly to $|\varphi_\uparrow\rangle$ in the Hilbert space norm $\||\psi\rangle\| := \sqrt{\langle\psi|\psi\rangle}$:

$$\lim_{k \to \infty} \||\varphi_k\rangle - |\varphi_\uparrow\rangle\| = 0,$$

but the branch-counting rule yields the constant sequence independent of k:

$$r_k = \frac{n_\uparrow}{n_\downarrow}$$

which does not converge to $r_\infty = 0$.

The result is damning of the branch-counting rule, but it may be that the difficulty goes deeper. It may be that the rule in this respect faithfully reflects Everett's concept of branching, and that it is not the rule that is deficient; the real problem may be with the concept of branching itself. This too may have no useful application to real measurement processes. The problem, in other words, may concern the Everett interpretation itself. After all, however singular the criterion for the existence of branches (that they have non-zero amplitude), it is *existence* that should count towards probabilities; existence is unequivocal, no matter the amplitude.[11] If the result is a probability interpretation that is at odds with the Born rule, and at best a singular function of the state, so much the worse for the Everett interpretation.

However it *is* a deficiency of the rule if *another* branch-counting rule can be found that does *not* suffer from this singular behaviour. The branches thus counted all unequivocally exist, whereupon probability, as it should, depends on their numbers. Is there one?

**5. Boltzmann's combinatorial definition of the equilibrium entropy**

We are looking for a method for counting branch states involving macroscopic numbers of degrees of freedom that

- depends non-trivially on the state
- is a continuous function of the state

and further, at least with respect to ratios in branch numbers, that

- is free of arbitrary convention.

---

[10] The same is true of many other kinds of measurements, but probably not of all – but we do not need so strong a claim.

[11] This has been denied by Vaidman (1998), (2019), who suggests that amplitude may be interpreted in terms of 'degree of existence'. (In effect, in what follows, we are showing that it may be interpreted in terms of 'numbers of existents'.)

Given which, the result may or may not be in agreement with the Born rule – and determine the fate of the Everett interpretation accordingly.

Substitute microstates in classical statistical mechanics for branch states in quantum mechanics, and the problem as posed has a ready precursor. A method for counting microstates was provided by Ludwig Boltzmann in one of his most important contributions to statistical mechanics, his 1879 memoir, in which he proposed that the entropy $S$ of a macrostate be identified with the logarithm of what Boltzmann called its thermodynamic probability $W$, by the equation:

$$S = k \log W.$$

More accurately, Boltzmann's idea was that entropy differences between states are identified with the logarithm of the ratio of the $W$'s, for those states: only ratios in $W$'s need be well defined, and thereby, only entropy differences.

This went hand-in-hand with another innovation, whereby two levels of description are needed: a fine-grained level of microstates (that Boltzmann called 'complexions') and a coarse-grained level of macrostates ('distributions'). The thermodynamic probability $W$ of a macrostate was to be determined as *the number of microstates energetically accessible to a gas, consistent with that macrostate* – it was a 'microstate-counting rule'. The method involved combinatorial techniques in the computation of these numbers, so it was often called 'the combinatorial approach' to the definition of entropy. The equilibrium state was then the most probable state – the one with the greatest number of accessible microstates.

Influential as it was, the question of whether the microstate-counting rule really defined probability, in classical statistical mechanics, never enjoyed wide consensus, and to this day remains controversial. But we need not address that question here; we are only interested in the definition of the rule -- on how Boltzmann defined his microstates, and with it his microstate-counting rule.

In classical theory, the actual state of a system of *N* particles is represented by a point in the *N*-particle phase space (or equivalently, *N* points in the 1-particle phase space). The number of accessible microstates, understood as phase-space points, consistent with a macrostate as a region of phase space, is then always the same, a continuum infinity. Boltzmann arrived instead at *finite* numbers by the following device: let 1-particle microstates be fine-grained cells in the 1-particle phase space of non-zero extension. Because of the bound on the total energy and volume, it follows that there are only finitely many microstates accessible to each of the *N* particles. But more specifically, he assumed that in the partitioning of the 1-particle phase space into accessible microstates, *all the microstates have the same extension $\varepsilon$ (the same 'volume' $\varepsilon$ in the one-particle phase space)*, as determined by Liouville measure. He *defined* the partitioning of the 1-particle phase space so that this was the case. The same then followed for the volume of N-particle microstates in the total phase space: each has the same volume $\varepsilon^N$. The number of accessible N-particle microstates thus defined, each of volume $\varepsilon^N$, contained in a macrostate, varies with the macrostate. Ratios in those numbers $W$ then give the relative probabilities of the macrostates, and $\varepsilon$ drops out as irrelevant; correspondingly, only entropy differences are defined, as independent of $\varepsilon$.

But the latter is not *quite* true. There is one final ingredient to Boltzmann's procedure: it is that macrostates may be defined by macroscopic thermodynamic quantities, like volume, temperature, and energy, independent of the fine-graining. In that case the procedure introduces rounding errors – the volume of the macrostate will not be an *exact* integral multiple of $\varepsilon$ – so, strictly speaking, we should take the limit $\varepsilon \to 0$, to obtain ratios that are exactly independent of $\varepsilon$.

All of this goes over to quantum mechanics, but with two important changes: first, we are interested in measurement processes, far from equilibrium; second, we are interested in temporal sequences of microstates, states now in Hilbert space. We also need the notions of coarse- and fine-grainings of regions of state space, now subspaces of Hilbert space. In short: we need the quantum histories formalism.

## 6. Quantum histories and quasiclassical domains

Consider first a single time. We fix a parameter space $\mathcal{D}$, for given degrees of freedom, and a partitioning $\alpha \in \mathcal{D}$ into disjoint microstates (for example, a fine-graining of configuration space). We may then define a correspondence $\alpha \to P_\alpha$ to a family of commuting projection operators on $\mathcal{H}$ such that for $\alpha \neq \alpha'$, the associated projectors are orthogonal and sum to unity:

$$P_\alpha P_{\alpha'} = 0; \quad \sum_\alpha P_\alpha = I.$$

For any state $|\varphi\rangle \in \mathcal{H}$, if

$$P_\alpha |\varphi\rangle = |\varphi\rangle$$

we say $|\varphi\rangle$ 'has property $\alpha$', and that the degrees of freedom in state $|\varphi\rangle$ 'take values in $\alpha$'. In parallel to Boltzmann's approach, we distinguish macrostates from microstates. The former, denote $\beta$, are unions of fine-grained microstates. The associated projectors are given by simple addition:

$$\sum_{\alpha;\, \alpha \subseteq \beta} P_\alpha = P_\beta. \tag{6.1}$$

A typical example for a model of measurement is where the macrostate $\beta$ specifies the state of the pointer-degrees of freedom as lying within a given latitude (as contained in a subspace of $\mathcal{H}$), integrating out (completely coarse-graining over) all other degrees of freedom.

To go from single to multiple times $t_1, t_2, \ldots, t_n$ we represent sequences of properties as n-tuples $\underline{\alpha}_n := \langle \alpha_1, \ldots \alpha_n \rangle$, where $\alpha_k$ takes the same range of values as does $\alpha$ in (6), save indexed to time $t_k$. In terms of these projection operators, the temporal nature of the ordering can be incorporated in quantum mechanics purely algebraically, in terms of *quantum history operators* $C_{\underline{\alpha}_n}$ *(of length n)*, also called 'chain operators':

$$C_{\underline{\alpha}_n} := P_{\alpha_n}(t_n) \ldots\ldots P_{\alpha_1}(t_1)$$

where $P_{\alpha_k}(t_k)$ is:

$$P_{\alpha_k}(t_k) := e^{iHt_k/\hbar} P_{\alpha_k} e^{-iHt_k/\hbar}$$

and *H* is the Hamiltonian. A space of all such histories (for given partitioning $\alpha \in \mathcal{D}$, Hamiltonian $H$, and coarse graining in time $t_1, \ldots, t_n$) is a *quantum history space*.

Quantum history operators do not in general commute with each other, but we can still define coarse-grained history operators, denote $C_{\underline{\beta}_n}$, by simple addition:

$$\sum_{\underline{\alpha}_n; \, \underline{\alpha}_n \subseteq \underline{\beta}_n} C_{\underline{\alpha}_n} = C_{\underline{\beta}_n} \qquad (6.2)$$

where $\underline{\alpha}_n \subseteq \underline{\beta}_n$ iff $\alpha_k \subseteq \beta_k$ for all $k = 1, \ldots, n$. Whether coarse- or fine-grained, these history operators are in the Heisenberg picture, where time-dependence is carried by operators rather than states. For the sake of familiarity, and to connect more cleanly with earlier material, we will continue to work in the Schrödinger picture, with the state changing in time. For this we need the operators:

$$C_{\underline{\alpha}_n}(t_n) := e^{-iHt_n/\hbar} C_{\underline{\alpha}_n}$$

(and similarly for coarse-grainings). Each $C_{\underline{\alpha}_n}(t_n)$ has the action: unitarily evolve the state to time $t_1$, and then project onto the component with property $\alpha_1$; unitarily evolve to time $t_2$, and project on to the component with property $\alpha_2$; and so on – to time $t_n$. It is the unfolding of a n-fold sequence of states, with decreasing norm, each state with a definite property, linked by unitary evolutions.

A history space is *decoherent* for a given initial state $|\varphi; 0\rangle$, if:

$$\underline{\alpha}_n \neq \underline{\alpha}'_n \implies \langle \varphi; 0 | C^\dagger_{\underline{\alpha}_n}(t_k) C_{\underline{\alpha}'_n}(t_k) | \varphi; 0 \rangle = 0.$$

Thus, the states $C_{\underline{\alpha}_n}(t_k)|\varphi; 0\rangle$ are orthogonal to one another, if the $\underline{\alpha}_n$'s differ at any time $t_1, \ldots, t_n$. This orthogonality condition ensures that these sequentially developing states over time – in the forward direction in time, with respect to $|\varphi; 0\rangle$ – do not interfere with each other; they do not recombine or cancel each other out. It also implies that there is only *one* way of arriving at each branch state $C_{\underline{\alpha}_n}(t_n)|\varphi; 0\rangle$ from the initial state $|\varphi; 0\rangle$, namely by passing through a sequence of states (vertical reading) having properties defined by the sequence $\underline{\alpha}_n$; any state that has evolved through any other sequence must be orthogonal to it. In this sense branch states, of the form $C_{\underline{\alpha}_n}(t_n)|\varphi; 0\rangle$, generalise Everett's notion of 'memory states'.[12]

The decoherence condition evidently brings with it an arrow to time. For example, if it is satisfied with respect to a partitioning $\alpha \in \mathcal{D}$ and state $|\varphi; 0\rangle$ for times $t_1, \ldots, t_n$, it will *not* be satisfied for the complex conjugate (time-reverse) of $|\varphi; t_k\rangle$. With respect to the latter, and for the same Hamiltonian, partitioning, and direction in time, the branches produced from $|\varphi; 0\rangle$ up to time $t_k$ will thereafter, with exquisite sensitivity to phase relations and amplitudes, all recombine. We are familiar with something very similar to this in connection with the second law of thermodynamics in classical statistical mechanics, with entropy production in place of branching. The two appear to be similar in the explanation of the ultimate origin of the asymmetry in time, in terms of a special initial state.

Decoherence alone is the most abstract of Everett's rules for branching, a generalisation of his notion of 'memory'. The stronger notion that we are looking for is where, in addition, the $\underline{\alpha}_n$'s satisfy definite equations – some set of equations $\mathcal{E}$, so that only those sequences that satisfy $\mathcal{E}$ (denote $\underline{\alpha}_n \vdash \mathcal{E}$) enter into the decomposition of the state as a superposition of branches. As a first stab, then, we are interested in a decoherent history space where the universal state can be written as a superposition of the form:

$$|\varphi; t_n\rangle = \sum_{\underline{\alpha}_n;\, \underline{\alpha}_n \vdash \mathcal{E}} C_{\underline{\alpha}_n}(t_n)|\varphi; 0\rangle. \tag{6.3}$$

---

[12] See Wallace (2012 pp.87-99) for a fuller discussion (in particular of the 'branching-decoherence' theorem).

This provides a vertical reading of the state in our previous sense: a superposition of sequences of states, each obeying a definite rule, as applies to fine-grainings of the state – the state up to a subspace of Hilbert space, at each time. But with equation (6.3), we (almost) have the condition for Gell-Mann and Hartle's concept of a quasiclassical domain.

## 7. The new branch-counting rule.

The branch-counting rule that we are looking for should not just be a count of quantum history *operator*s, but should depend explicitly on the state. The obvious rule is to count the number of non-zero *branches*, by the end of an experiment, at time $t_n$; that is, the number of non-zero branch states at $t_n$ of the form:

$$C_{\underline{\alpha}_n}(t_n)|\varphi; 0\rangle \neq 0.$$

Equivalently, it is to count the number of fine-grained history operators of length $n$ that satisfy:

$$\|C_{\underline{\alpha}_n}(t_n)|\varphi; 0\rangle\| > 0. \tag{7.1}$$

On the conventional interpretation of quantum histories, (9) is just the condition that histories have non-zero probability. But we do not need this interpretation; we speak simply of branches of non-zero amplitude.

The rule may be obvious, but it is wrong. It leads to the same singular behaviour under variations of the quantum state that we encountered earlier. To give another example, consider quantum Brownian motion, where the parameter space $\mathcal{D} = \mathcal{D}_S \times \mathcal{D}_E$ is the Cartesian product of the configuration space $\mathcal{D}_S$ of the centre of mass degrees of freedom *S*, with a parameter space $\mathcal{D}_E$ (that we coarse-grain over completely) of a large number of much lighter degrees of freedom (the scattering environment *E*) – for the details, see Joos et al (2002 §5.1). The fine-grained histories $\underline{\alpha}_n$ of S, including recoil from scattering processes, are trajectories in $\mathcal{D}_S$, with each $\alpha_k$ a fine-grained cell in $\mathcal{D}_S$ at time $t_k$. Now suppose the initial state $|\varphi; 0\rangle$ is non-zero only in some coarse-grained region $\beta \subset \mathcal{D}_S$; then from the kinematic part of the Hamiltonian, for *any* $t > t_0$, the state $|\varphi; t\rangle$ that (continuously) evolves from $|\varphi; 0\rangle$ is non-zero throughout all of $\mathcal{D}_S$; for any $t > t_0$, the condition (9) is satisfied for every $\underline{\alpha}_n$. Numbers counted in this way are discontinuous functions of the state.

What this example also shows is that the definition of quasiclassical domain should not call for *exact* equality in equation (6.3), as Gell-Mann and Hartle made clear. In their words, it is a decoherent history space 'with probabilities peaked on quasiclassical histories'. Again, we say rather with *amplitudes* peaked on quasiclassical histories, for we do not yet have a probability interpretation. The vertical reading of the state provided by a quasiclassical domain is rather that *to a very good approximation* it is a superposition of rule-bound branches, where the approximation is as ever controlled by the Hilbert-space norm. Identity in equation (6.3) is replaced by approximate equality.

That seems just as it should be. Models in decoherence theory invariably involve integrating out 'irrelevant' ('fast') degrees of freedom (usually the environment), whose contributions to the dynamics of the 'relevant' ('slow') degrees of freedom are only approximately taken into account in the equations $\mathcal{E}$. The universal state evolving under the Schrödinger equation knows no such limitation. Likewise, the equations $\mathcal{E}$ are only defined up to a given approximation or idealisation; so too is the decoherence condition, in realistic models, where the history space is not fine-tuned to the state. All these approximations are defined in terms of continuity in the Hilbert space norm. It follows that Everett's branches are not *exactly* defined, nor are they *exactly* autonomous from one another -- there remain tiny interference effects, infinitesimal ripplings in amplitudes. Again, no probabilistic concepts are being smuggled in.[13]

However, we are no nearer to a branch-counting rule that is a continuous function of the state. Equation (7.1) is clearly a necessary condition on branches, but just as clearly it is insufficient. We propose instead to use the analogue of Boltzmann's method: we fix a unit $\tau \ll \||\varphi; 0\rangle\|$ in the Hilbert space norm and *define* the fine-grained partitioning $\underline{\alpha}_n$ so that:

$$\left\|C_{\underline{\alpha}_n}(t_n)|\varphi; 0\rangle\right\| = \tau. \qquad (7.2)$$

The relative probability of disjoint coarse-grained histories $\underline{\beta}_n$, $\underline{\beta}'_n$ (which may be events at some single time $t_k$, completely coarse-graining all other times), is then the ratio of the numbers $N_{\underline{\beta}_n}$, $N_{\underline{\beta}'_n}$ of fine-grained histories into which $\underline{\beta}_n$, $\underline{\beta}'_n$ are partitioned, satisfying (10).

To see what those ratios are, observe that from the approximate orthogonality condition ('medium decoherence' in Gell-Mann and Hartle's terminology), and from the definition of coarse-graining equations (6.1) and (6.2):

$$\sum_{\underline{\alpha}_n \subseteq \underline{\beta}_n} \left\|C_{\underline{\alpha}_n}(t_n)|\varphi; 0\rangle\right\|^2 \approx \left\|\sum_{\underline{\alpha}_n \subseteq \underline{\beta}_n} C_{\underline{\alpha}_n}(t_k)|\varphi; 0\rangle\right\|^2 = \left\|C_{\underline{\beta}_n}(t_k)|\varphi; 0\rangle\right\|^2. \qquad (7.3)$$

From the condition on the fine-graining $\underline{\alpha}_n$, equation (7.2), each summand on the LHS of (7.3) is just $\tau^2$. Let there be $N_{\underline{\beta}_n}$ such, and similarly $N_{\underline{\beta}'_n}$ for the coarse-graining $\underline{\beta}'_n$; it then follows:

$$\frac{N_{\underline{\beta}_n}}{N_{\underline{\beta}'_n}} = \frac{N_{\underline{\beta}_n}\tau^2}{N_{\underline{\beta}'_n}\tau^2} \approx \frac{\left\|C_{\underline{\beta}_n}(t_n)|\varphi; 0\rangle\right\|^2}{\left\|C_{\underline{\beta}'_n}(t_n)|\varphi; 0\rangle\right\|^2}. \qquad (7.4)$$

The right-hand side is the ratio of the Born rule quantities. Choosing $\underline{\beta}'_n$ as the identity ($\beta_k = \mathbb{I}$ for $k = 1, \ldots, n$), the number $N_{\underline{\beta}'_n}$ is just the total number of branches satisfying (7.2); thus normalised, we obtain the Born rule probability of $\underline{\beta}_n$ for initial state $|\varphi; 0\rangle$:

---

[13] The objection has often been made that decoherence theory cannot be used to interpret the quantum state in terms of probability, because it presupposes the concept of probability (see e.g. Baker (2007), Dawid and Thébault (2015), and for a reply, Wallace (2012 p.253-54), Saunders (2021)); I am trying to pre-empt that criticism.

$$\Pr\left(\underline{\beta}_n\right) = \frac{\left\|C_{\underline{\beta}_n}(t_n)|\varphi;0\rangle\right\|^2}{\||\varphi;0\rangle\|^2}.$$

The new branch-counting rule therefore provides an excellent approximation to the Born rule. In fact, so long as the $\underline{\beta}_n$'s are defined by (6), (7), the only approximation involved concerns orthogonality. But as with Boltzmann's procedure, we should allow that the coarse-graining $\beta_k$ at each time $t_k$ may be defined without reference to any fine-graining, in which case the second equality in equation (7.3) is approximate too: $\underline{\beta}_n$ will not *exactly* be the union of an integral number of fine-grained histories $\underline{\alpha}_n$ satisfying (7.2); there will be rounding errors, bounded by $\tau$. Realistically, however, $\tau$ can be chosen as *extremely* small, consistent with the decoherence condition, as witness the characteristic lengths and times of branching in thermal environments.

How does the new rule behave, in terms of its dependence on the state? Consider, for fixed $\tau$, a sequence of states $|\varphi_k;0\rangle$ converging to $|\varphi;0\rangle$ in the norm topology, as before:

$$\lim_{k\to\infty}\||\varphi_k;0\rangle - |\varphi;0\rangle\| = 0. \qquad (7.5)$$

Let the ratio as given by the RHS of (7.3) for the state $|\varphi_k;0\rangle$ be the real number $r_k$, and for $|\varphi;0\rangle$ let it be $r$. From (7.5) we know $r_k \to r$. Let the ratio of branch numbers for the state $|\varphi_k;0\rangle$ be $p_k$, and for the state $|\varphi;0\rangle$ be $p$. Then $|p_k - r_k| < \tau$, and $|p - r| < \tau$. Since $r_k$ converges to $r$:

$$\lim_{k\to\infty}|p_k - p| < 2\tau. \qquad (7.6)$$

We cannot without further ado take the limit $\tau \to 0$, for the reason already stated: the fine-graining $\underline{\alpha}_n$, of length $n$, cannot be taken to zero, on pain of violating the decoherence condition. So for the moment (7.6) gives us something slightly less than desired: a branch-counting rule that is only approximately continuous with the state, bounded by $\tau$. We shall see how to do better in a moment.

## 8. Discussion

An obvious objection to the new branch-counting rule is that it is question-begging: according to the new rule, put in terms of probability, equi-amplitude branches are equiprobable; therefore probability is assumed at the beginning.

There is something to this objection, but it misses an important point. We have provided a rational as to *why* probability should depend on amplitude, and not on phase – and not, for example, on the number of four-leaf clovers in each branch. Amplitude dictates the structure of the quantum state, and with it branching structure; and relative numbers of branches, probabilities. (The old branch-counting rule offered this rational too, but on the basis of a singular structure to the state that is never actually realised.) As for whether equiprobability must be assumed, see below.

A variant of the same worry is that the new rule is ad hoc, expressly designed to yield the right ratios, in accordance with the Born rule. Yet that would seem unfounded given that the method is the same as Boltzmann's (we consider whether they really are the same in the next section). Moreover, the method appears to be a natural one: we learn from decoherence theory that branching is pervasive, and that there are countless non-zero decohering states, of ill-defined number, varying discontinuously with the state; we group together similar decohering states, so that each group (superposition) has the same 'size' in state-space -- the same Hilbert-space norm – and count those groups; as in classical statistical mechanics, we group similar states, so that each group has the same 'size' in state-space, the same Liouville measure – and count those groups. In both cases we obtain ratios in numbers that covary with the state. There may be no true number of the relevant microstates, in each case, but there may yet be true ratios.

Another objection is that the notion of approximation itself involves probability, as with frequentism in the philosophy of probability. The most one can say (in a one-world setting) of the actual relative frequency of an outcome (involving only finitely-many runs of an experiment) is not that it is the probability of that outcome on each trial, nor that it is an approximation to that probability (these the main difficulties for naïve frequentism), but only that it is *probably* that probability, to a given approximation (this the law of large numbers). As such (the objection continues) the new branch-counting rule, as a version of naive frequentism, can do no better than Everett's law of large numbers.

This objection is unfounded, however. The nature of the approximation is *not* probabilistic – it is not that with high amplitude the branch-counting rule is approximately as given by the Born rule quantity, it is that with certainty it is given approximately by the Born rule quantity. The new branch-counting rule is about all the branches produced on a single measurement, not a privileged subset; ratios in their numbers invariably approximate the Born rule quantities, differing by $\tau$ at most. The new rule is like naïve frequentism, as concerning the count of actual outcomes, but instead of being spread over different times in a single world, it is spread over worlds at a single time. The difference, in providing a matching of relative frequencies to probabilities, is decisive. Of course, unlike actual relative frequencies in a single world, following multiple measurements, the relative frequencies across worlds (for a single measurement) are unobservable. Branch-counting provides a new story about what probability is, not a new method of measurement.

At this point it is worth reprising another argument against the old branch-counting rule, due to Wallace (2012 p.120). Suppose we consider a measurement of spin, at time $t_1$; then by the old branch-counting rule there are two outcomes, each with probability one half. But suppose a second measurement is made in the spin-up branch, but not the other, at time $t_2 > t_1$; now there are two branches with the spin-up result, and only one with spin-down; so the probability is two-thirds. The probability for the event at $t_1$, as given by branch-counting, depends on the various measurements performed at later times $t_2$, and so on for all subsequent times.[14]

This being so, why did we need the argument from continuity in the state? Because there is a variant of the old rule that does not depend on the lengths of branches considered at subsequent times. Call it 'the equi-outcome rule': every outcome in an experiment has equal probability, so long as it has non-zero amplitude. This respects the intuition that existence is what matters, regardless of amplitude, immediately following a measurement.

---

[14] The parallel with the Sleeping Beauty problem is obvious and has been used (for example by Lewis (2009)) to suggest that instabilities in probabilities, if defined in terms of self-locating uncertainty, are inevitable in the Everett interpretation.  See also §8

The difference is that probabilities of temporal sequences of outcomes, for various experiments, are computed in the usual way, using the equi-outcome rule for each experiment instead of the Born rule. Branches now have unequal probabilities, as depending on their histories of branching. (In the case just given, the two branches with spin-up have probabilities one quarter, the one with spin-down one half). But the equi-outcome rule falls to the same argument from continuity as before. It fails, when applied to realistic experiments, to covary with the state; it is not an interpretation of the state. [15]

The old branch-counting rule thus came in two conflicting versions, depending on the lengths of the branches, does the new branch-counting rule too? Consider branches of much greater length $N$, where $t_N \gg t_n$, possibly following many other kinds of measurements. The question is whether the ratio of the numbers of those length N branches, containing $\beta_n$, $\beta'_n$, at time $t_n$, all of equal amplitude, are the same as for branches of length n, all of equal amplitude.

A little reflection shows that the only difference is that the longer the branch, the greater the freedom to use much smaller values for $\tau$, without jeopardy to the decoherence condition. The number of branches is correspondingly increased, not because they differ from one another in ever more microscopic intricacy, at time $t_n$, but because they differ in similarly intricate ways at future times. The ratios for branches of length N differ from those of length n, in that rounding errors, already tiny, are further reduced. The limit $\tau \to 0$ is the limit $t_N \to \infty$, and in that limit in full rigor, the new rule is a continuous functional of the initial state. (Here N is not the number of runs of the same experiment, as with one-world frequentism; the experiment may be performed just the once. N is the number of time increments, not the number of trails.)

What of the fact that the new counting rule involves ratios in the *squares* of the amplitudes? It derives, like the continuity requirement, from the norm on Hilbert space. For the space of square-integrable functions $\phi: X \to \mathbb{C}$ for some measure space $(X, \mu)$, the norm is:

$$\|\phi\| = \sqrt{\int_X |\phi|^2 d\mu}.$$

This was in play in equation (7.3); if instead we had considered an $L^p$ space (the space of $p$-integrable functions on $(X, \mu)$), with norm

$$\|\phi\|_p = \sqrt[p]{\int_X |\phi|^p d\mu}$$

we would have obtained not the Born rule, but a $p^{th}$-power rule. So did we put in the Born rule by hand, in choosing $p = 2$?[16] But an $L^p$ space, although a topological linear space (a 'Banach space'), is only an inner product space for $p = 2$; only in an inner product space is the notion of orthogonality defined. Tamper with quantum mechanics at this level, and you risk talking about a different theory altogether.

---

[15] It has recently been shown to permit super-luminal signalling as well (McQueen and Vaidman (2019); not so the new rule.

[16] An objection due to Mateus Araújo (personal communication, 18/6/18)); I had earlier noted this possibility (Saunders (2004)) in the context of a derivation akin to Deutsch (1999).

There remains the question of how the new rule bears on rational agency. It is an important virtue of branch counting that it is independent of degrees of belief, personal identity, and uncertainty; independent of whether there are any rational agents, or observers, at all. But it surely *should* matter to agency, where there are observers cognizant of quantum mechanics and the new branch-counting rule. It is entirely plausible, prior to branching, that we *ought t*o care more about outcomes that happen in many branches, than we care about outcomes that happen in comparatively few; and similarly, that we ought to expect to see those outcomes, that are contained in the overwhelming majority of branches, rather than outcomes that are contained in comparatively few.

The normative dimension to the Born rule in the Everett interpretation has in recent literature been explored in great detail, most notably in Saunders et al (2010) and Wallace (2012). On the table is a demonstration that an agent, if rational, in acting in accordance with certain axioms, and if cognisant of the unadorned unitary formalism of quantum mechanics, should choose among quantum games as if ranking them by their expected utilities, using the Born rule. Those axioms have been much questioned, but to inconclusive effect (for a recent survey see Vaidman (2021)); here we are mainly concerned with their relation to the new branch-counting rule, rather than their truth.

At the heart of Wallace's proof (and the core of the original ideas in Deutsch (1999)) is a demonstration than when experimental outcomes have equal amplitudes, an agent should be indifferent as to which outcome has which reward. That translates into the condition: when outcomes have equal numbers of branches, counted using the new rule, agents should be indifferent as to which outcome has which reward. But that just *is* to use the branch-counting rule. It follows directly that agents who are rational, in Wallace's sense, should be guided by ratios in branch numbers, in ours.

The only question, therefore, is whether Wallace's axioms are available to branch-counters. The answer is not entirely obvious, not least because among them is 'branching indifference' - an axiom that has often been justified (including, at times, by Wallace) both on the basis that branch numbers are undefined, and because branching is pervasive (therefore 'do not care about branching per se', Wallace (2012 p.170)). On branching as pervasive we can agree; but on branch numbers?

An inspection of his definitions shows that by 'branch', Wallace meant 'non-zero branch state'. We can agree with him not to care on the number of *those*. But more revealing is the axiom itself, that if an operation leaves the future selves of an agent in M different macrostates, all with exactly the same reward, the agent should be indifferent as to whether or not the operation is performed. For branch-counters, this axiom translates into: an agent should be indifferent as to whether the number of branches, all with the same reward, are collected together into M coarse-grained branches, each macroscopically distinct, without change of the reward. But that would appear to be an eminently reasonable principle: why care about grouping branch numbers in these various ways, so they are noticeably different, when they all have the same reward? Especially so, if all we care about are the rewards, the usual assumption in operational approaches to rational choice theory.

There remains one other source of difficulty, in taking over Wallace's proof, for it depends on his axiom of 'problem continuity', which is similar to our requirement of continuity in the state. As a matter of mathematical rigour, his proof requires that we take the limit of branches of arbitrarily long length. But with that all his axioms (and in particular 'branching indifference') are available to branch-counters, who if anything will find them more compelling. Moreover, they (possibly simplified) are still needed, if wanted is a normative dimension to branch-counting. On the other hand, branch-counting does better in providing explanations. Do we explain the observed statistics, by demonstrating that it is the statistics a rational agent, cognisant of quantum mechanics, ought to bet on? Better, we explain the observed statistics, by showing that they are overwhelmingly more common among all the worlds. Digging deeper, we see that this involves treating branches as equiprobable, and we may well ask with what rationale -- at which point Deutsch's symmetry argument applies, and we have the normative argument as well.[17]

This being so, much recent debate on the meaning of probability in the Everett interpretation seems misplaced. We have just dealt with the objection, for example in Albert (2016), that betting behaviour cannot be explanatory of the observed statistics. There is also the radical proposal, proposed by Deutsch (2016) and Brown and Porath (2020), that physical probability simply has no meaning in the Everett interpretation, and there are only agent-specific personal probabilities. But ratios in branch numbers are physical, and in view of the similarities with naïve frequentism, to deny that they have anything to do with probability seems capricious. Wallace argues rather that physical probability just is whatever rational agents ought to track in assigning personal probabilities, and since they ought to track the squares of the branch amplitudes, the latter are physical probabilities. This argument has a considerable background in philosophy, however, and it can be rendered more simply as rational agents ought to treat equi-amplitude branches as equiprobable, whereupon ratios in branch numbers are physical probabilities. Meanwhile the debate over the nature of persons and first-person knowledge, and probabilities in the sense of self-locating probabilities (Saunders 2010, Wallace 2012, Sebens and Carroll 2018, McQueen and Vaidman 2019), whatever its intrinsic interest, can be set to one side, as not needed for the Everett interpretation; if facts about self-location are irrelevant to empirical tests, as argued by Adlam (2014), that may be just as well.

## 9. The origins of quantum mechanics

The final objection to the new branch-counting rule that we consider relates to the parallel with Boltzmann's method. Is our method and his really the same? If not, the new rule is open to the charge that it is ad hoc.

I shall approach this question obliquely, noting that Boltzmann's method for counting microstates introduced in 1879 led, some two decades later, to the discovery in 1900 of Planck's constant – to the discovery, by Planck, that what from the point of view of the Liouville measure of certain macrostates, are rounding errors, introduced by a choice of unit of phase space volume $\varepsilon$, are actually essential to getting the right equilibrium distribution (as determined by experiment). This discovery led to quantum mechanics.

The system studied by Planck was black-body radiation, and specifically its energy density as a function of temperature and frequency. It was known (from the Wien displacement law) to involve an unknown constant with the dimensions of action. The choice of $\varepsilon$ on what was eventually recognised, by Bose, as the one-photon phase space, required to give the 'right' rounding errors, was that constant: Planck's constant $h$. With that it followed not just that *ratios* of numbers of microstates are independent of convention, but those numbers themselves.

---

[17] My thanks to David Deutsch on this point (personal communication, 7/7/21).

That raises two questions. Does it show that Boltzmann's method only really made sense, given this ultimate, discrete structure to phase space? – with the implication that unless something similar is forthcoming for Hilbert space, our use of his method is not safe. Second and relatedly, does it show that Boltzmann's method in fixing a unit $\varepsilon$ wasn't really about the state at all, but about the structure of the state space?

The latter worry first. Planck had hit on the empirically correct black-body formula more or less by guesswork; and was then led by a process of reverse engineering to a combinatorial expression, the logarithm of which gave the right entropy as implied by his formula. In this way he obtained, for radiation in spatial volume *V* with frequency $\nu_s$ (where s is a discretisation of frequency) the expression:

$$W = \prod_s W_s = \prod_s \frac{(z_s + n_s - 1)!}{(z_s - 1)!\, n_s!} . \qquad (9.1)$$

Here $z_s$ was the number of what Planck called 'oscillators', and $n_s$ the number of 'energy quanta' of energy $h\nu_s$. Maximising (9.1), on variation of the numbers $n_s$, subject to fixed total energy (using the method of Lagrange undetermined multipliers), and determining $z_s$ as the number of standing waves for volume V and frequency $\nu_s$ (following Jeans), Planck arrived at the right spectral energy density as a function of temperature and frequency -- the Planck black-body formula. The remaining difficulty was to give a rational for (9.1), and specifically, the numbers $W_s$.

It took almost a quarter of a century to understand this combinatorics expression correctly: $W_s$ is the number of available microstates for $n_s$ photons distributed over $z_s$ one-particle states, in frequency interval indexed by s, all of the same 'elementary volume' $h$ in the one-photon phase space, without regard for which photon is in which microstate (the latter since photons are 'indistinguishable particles'). When it was eventually understood in this way, by Bose and Einstein in a series of papers in 1924-25, it was immediate how to apply the same technique to a gas of material particles, replacing photons by non-relativistic molecules. The same combinatorial expression (9.1) applied, but where s labels intervals of momentum, rather than frequency, and with an additional constraint on total particle number, as follows from conservation of mass. Thus was born the theory of what is now called the Bose-Einstein gas.

This fine-graining of classical phase space in terms of $\varepsilon = h$ involved a measurable (and universal) constant. Nothing like this seems to be in play in our use of a unit $\tau$ in the Hilbert space norm. But to see the connection, we need to understand Bose and Einstein's procedure in purely quantum mechanical terms. This was soon done, in Dirac's treatment of the new gas theory in 1926, whereupon each term in the product (9.1) was newly interpreted: $W_s$ was now the dimensionality of the symmetrised subspace of the Hilbert space of $n_s$ particles, of frequency $\nu_s$, confined to spatial volume *V*, where the Hilbert space of each particle has dimension $z_s$.[18]

---

[18] Jeans' standing waves were in effect orthogonal states spanning the one-photon Hilbert space for given degrees of freedom (a given frequency interval). For a recent history of these discoveries, see Saunders (2020).

The counting method thus seems to have nothing to do with the state at all, but only with the structure of (symmetrised subspaces of) Hilbert space, quite contrary to what we have been doing, where the number depended on the state. But that is the wrong way of looking at it. Rather, Boltzmann, and Planck, and Einstein, and Dirac, all in effect used the *equilibrium* state – in the quantum case, a density matrix entirely degenerate with respect to each of these symmetrised subspaces, essentially giving equal 'volume' to each orthogonal dimension in each frequency range – this the same as the count of the dimensionality of each subspace. The procedure can in turn be redescribed as: for each frequency range, count the number of equi-amplitude fine-grained branches, given a totally degenerate density matrix, where the fine-graining goes all the way down to one-dimensional projection operators. This gives the number of dimensions of the subspace of Hilbert space for each frequency range, the numbers $W_s$, but as an exceptional case: it is possible to fine-grain in this way, whilst still satisfying the decoherence condition, because the latter is exactly satisfied for any fine-graining, given a totally degenerate density matrix for each frequency range – as required of an equilibrium state. In going over to non-equilibrium processes, like measurements, we should replace equilibrium states by non-equilibrium states, and worry about the decoherence condition, and equations for branches – so we do *not* go all the way down to one-dimensional projectors. This was just our procedure.

As for the suggestion that it is only because an absolute significance was found for Boltzmann's fine-graining (as fixed by Planck's constant) that the method had any validity – so that only given a lower limit on $\varepsilon$, did it make sense to count in Boltzmann's way - it would be hard to imagine a greater inversion of the rationale that was historically offered. It was only insofar as ratios in numbers of microstates W were approximately independent of $\varepsilon$, that the method was judged to be sound. Planck was repeatedly criticised for failing to take the limit $\varepsilon = h \to 0$. That the method led to an experimental test of this proposition is to its extraordinary credit; conceivably, it may one day lead to a test of the use of a continuous norm on Hilbert space. But its legitimacy should not depend on it. It would be perverse to require of a method that was revolutionary in one context, that applied in another, it prove revolutionary again.

**Acknowledgments**

My thanks to Harvey Brown, David Deutsch, Paul Tappenden, Lev Vaidman, and David Wallace, for helpful comments, encouragement, and criticism.